\documentclass[11pt]{article}
\usepackage[colorlinks, linkcolor={blue},citecolor={red}]{hyperref}
\usepackage{amsmath,amsfonts,amssymb,multicol,bm,nccmath,bbm}
\usepackage{graphicx}
\usepackage{epstopdf}
\newcommand{\nc}{\newcommand}
\nc{\ba}{\begin{eqnarray}} \nc{\ea}{\end{eqnarray}}
\newcommand\be{\begin{equation}}
\newcommand\ee{\end{equation}}
\nc{\D}{\overline{\mbox{D3}}}

\nc{\ga}{\gamma} \nc{\tnu}{\tilde{\nu}} \nc{\tmu}{\tilde{\mu}}

\nc{\x}{{\bf{x}}}

\input{epsf.sty}
\begin{document}
\title{{\bf Ghost-free higher order gravity from  bi-gravity}}

\author{Nima Khosravi\thanks{%
		e-mail: n-khosravi@sbu.ac.ir}\,\, and Reza Rajaee\thanks{%
		e-mail: r.rajayei@mail.sbu.ac.ir }
	\\\\
	{\small {\it Department of Physics, Shahid Beheshti University, G.C., Evin, Tehran 19839, Iran}}
}

\maketitle
\begin{abstract}
In this work we studied the higher order gravity model which corresponds to Hassan-Rosen ghost-free bi-gravity. To do this we absorb one of the metrics in bi-gravity model in favor of the other metric in a recursive way. For the second (recursion) step	we get conformal gravity same as what has been done in the literature. We generalize this idea by calculating up to the fourth order gravity. To reach to a ghost-free higher order gravity we need to go to infinite-derivative order gravity but we can see our model as an effective theory. So adding higher order terms results in wider range of validity of our model. We emphasize that graviton mass controls the validity of perturbative approach.

\end{abstract}

\section{Introduction}
The Einstein-Hilbert action (EH) is a unique way to model gravitational force by introducing a massless spin-0 particle named graviton. However there are many attempts to generalize EH due to different reasons. One reason is explaining the observational data especially the late time acceleration and its tensions with the standard EH. The other one is the theoretical concerns especially the cosmological constant problem. On the other hand there is a curiosity that if there is another consistent model describing the gravitational force or not? A very interesting generalization of EH is trying to see if there is a covariant Lagrangian for a massive spin-2 particle. It has been shown that this model exist which is now famous as dRGT massive gravity \cite{deRham:2010kj,Hassan:2011zd} and has been shown to be ghost-free by Hamiltonian analysis \cite{Hassan:2011hr,Hassan:2011ea,Hassan:2011tf} for both massive and bi-gravity, for a review see \cite{deRham:2014zqa,Schmidt-May:2015vnx}. This field of research has been considered from different viewpoints: spherically symmetric solutions have been studied extensively e.g. in \cite{Koyama:2011xz} and cosmological solutions in e.g. \cite{DAmico:2011eto,Gumrukcuoglu:2011ew,Khosravi:2011zi}.

However an important question in massive gravity and bi-gravity is if they can answer the cosmological constant problem. Basically the idea is built on the 'tHooft conjecture that says a small parameter can be natural if a (quantum) symmetry exists when we set that parameter to vanish. Partially massless (PM) gravity is an attempt to look for such a symmetry in the field of massive gravity \cite{deRham:2013wv} and bi-gravity \cite{Hassan:2012gz}. In bi-gravity scenarion it has been shown that PM is related to higher order gravity models and specially conformal gravity \cite{Hassan:2013pca}. The approach in \cite{Hassan:2013pca} is as follow: one can use the equation of motion for one of the metrics to eliminate one of the metrics in favor of the other one. This step cannot be done exactly and can just be realized recursively. As we mentioned it has been shown that up to second step one get conformal gravity which belongs to the second order gravity family. However the main issue in this approach is when we cut the recursion procedure since it results in ghost in corresponding higher order gravity model as it is well-known e.g. for conformal gravity. On the other hand since we know the bi-gravity model is ghost free then we expect to have a ghost free higher order gravity if we keep all the infinite terms in recursion procedure. Our approach which is based on this idea is generalization of \cite{Hassan:2013pca} for the next two orders i.e. fourth order gravity. This is useful if we look at this procedure as an effective theory: if graviton mass be small then we can trust to our perturbation theory and even finite terms of higher order Riemann tensor terms will be okay under a cut-off energy scale. However recently it has been claimed that the higher order gravity from bi-gravity is ghost free \cite{Akagi:2018osc}. On the other hand higher order gravity models has been considered extensively in the literature including infinite-derivative gravity models \cite{Talaganis:2014ida}. Quantum gravity (e.g. string theory) suggests to have higher order gravity models as UV corrections to standard Einstein-Hilbert action. Again if one is far from Planck energy scale regime then can trust to effective theory approach and work with few first terms of the expansion.

In the next section we study the mathematical setup of our idea by giving a quick review on \cite{Hassan:2013pca}. Then we generalize their results up to fourth order. Finally we conclude and suggest some ideas for the future.

\section{Mathematical Setup}

We start from bi-gravity action 

\begin{eqnarray}\label{bi-gravity-action}
S=m^2_g\int d^4x\sqrt{-\det g}R^{(g)}+m^2_f\int d^4x\sqrt{-\det f}R^{(f)}-2m^2_gM^2\int d^4x\sqrt{-\det g}\sum_{n=0}^{4}\beta_ne_n(S)
\end{eqnarray}
where the first two terms are kinetic terms for $g_{\mu\nu}$ and $f_{\mu\nu}$ respectively. The third term is the famous dRGT mass term where $ S^{\mu}_{\nu}\equiv(\sqrt{g^{-1}f})^{\mu}_{\nu}$ and $e_n$'s are elementary symmetrical polynomials. The corresponding equations of motion with respect to $g_{\mu\nu}$ and $f_{\mu\nu}$ are respectively
\begin{align}
&R^{(g)}_{\mu\nu}-\frac{1}{2}g_{\mu\nu}R^{(g)}+2M^2g_{\mu\rho}\sum_{n=0}^{3}(-1)^n\beta_n(Y_{(n)})^\rho_\nu(S)=0\label{eqn-g}\\
&R^{(f)}_{\mu\nu}-\frac{1}{2}f_{\mu\nu}R^{(f)}+2\alpha^2M^2f_{\mu\rho}\sum_{n=0}^{3}(-1)^n\beta_{4-n}(Y_{(n)})^\rho_\nu(S^{-1})=0\label{eqn-f}
\end{align}
where
\begin{align}
&(Y_{(n)})^\rho_\nu(S)\equiv\sum_{k=0}^{n}(-1)^ke_k(S)(S^{n-k})^\rho_\nu.
\end{align}
It has been shown that one can write the above equation of motion (\ref{eqn-g}) as 
\begin{eqnarray}\label{PvsS}
P^{\mu}_{\nu}-[P]\delta^{\mu}_{\nu}=-M^2\Big [&&\beta_0e_0(S)\delta^{\mu}_{\nu}-\beta_1\Big (e_0(S)S^{\mu}_{\nu}-e_1(S)\delta^{\mu}_{\nu}\Big )\nonumber\\
&&+\beta_2\Big (e_0(S)(S^2)^{\mu}_{\nu}-e_1(S)S^{\mu}_{\nu}+e_2(S)\delta^{\mu}_{\nu}\Big )\nonumber\\
&&-\beta_3\Big (e_0(S)(S^3)^{\mu}_{\nu}-e_1(S)(S^2)^{\mu}_{\nu}+e_2(S)(S)^{\mu}_{\nu}-e_3(S)\delta^{\mu}_{\nu}\Big )\Big ]
\end{eqnarray} 
where we have used Schouten tensor $P_{\mu\nu}\equiv R_{\mu\nu}^{(g)}-\frac{1}{6}g_{\mu\nu}R^{(g)}$ and defined $[P]=g^{\mu\nu}P_{\mu\nu}$. The main idea is based on the solution of the above equation for $S$ as a function of $P$ i.e. $S^{\mu}_{\nu}=S^{\mu}_{\nu}(P)$. It seems it is impossible to solve the above equation exactly but it is possible if we do it recursively. For the moment let's assume we have $S$ as a function of $P$ which is given by $g_{\mu\nu}$ and its derivatives. Then we can read $f_{\mu\nu}$ from definition of $S$ as $f_{\mu\nu}=g_{\mu\rho}(S^2)^{\rho}_{\nu}$ which means $f_{\mu\nu}$ is given by $g_{\mu\nu}$. Now we can replace all $f_{\mu\nu}$'s in bi-gravity action (\ref{bi-gravity-action}) which results in a higher order gravity action for $g_{\mu\nu}$. The plugging procedure has two parts which are straightforward: i) For the kinetic term (i.e. second term in (\ref{bi-gravity-action})) we calculate  determinant of $f_{\mu\nu}$ by using $\sqrt{-\det f}= e_4(S)\,\sqrt{-\det g}$ which should be multiplied to its corresponding Ricci scalar
\begin{align}
R_{\mu\nu}^{(f)}=R_{\mu\nu}^{(g)}+2\triangledown_{[\mu}C_{\sigma]\nu}^{\sigma}-2C_{\nu[\mu}^{\rho}C_{\sigma]\rho}^{\sigma}
\end{align} 
where
\begin{equation}
C^{\sigma}_{\mu\nu}\equiv\frac{1}{2}(f^{-1})^{\sigma\rho}\Big (\triangledown_{\mu}f_{\nu\rho}+\triangledown_{\nu}f_{\mu\rho}-\triangledown_{\rho}f_{\mu\nu}\Big ).
\end{equation}
ii) In addition since the potential term (i.e. third term in (\ref{bi-gravity-action})) is a function of $S$ so we can easily convert it to a function of $g_{\mu\nu}$ and its derivatives. In the case of partially massless (PM) it has been shown \cite{Hassan:2012gz} that the free parameters in (\ref{bi-gravity-action}) should be as follow
\begin{equation}\label{PM-param}
\alpha^2\beta_0=3\beta_2,\qquad\qquad\qquad 3\alpha^2\beta_2=\beta_4,\qquad\qquad\qquad \beta_1=\beta_3=0,
\end{equation}
where	
$ \alpha\equiv\frac{m_f}{m_g} $.

In the follwing subsections we will solve $S$ as a function of $P$ recursively and calculate the final higher order Lagrangian for a single metric $g_{\mu\nu}$. 

\subsection{Quadratic Order}
For the first step, to solve (\ref{PvsS}) we can assume $S$ is written as 
\begin{align}\label{HD-2}
S^{\mu}_{\nu}=a\delta^{\mu}_{\nu}+\frac{1}{M^2}\Big (b_1P^{\mu}_{\nu}+b_2[P]\delta^{\mu}_{\nu}\Big )+\frac{1}{M^4}\Big (c_1(P^2)^{\mu }_{\nu}+c_2[P]P^{\mu}_{\nu}+c_3[P^2]\delta^{\mu}_{\nu}+c_4[P]^2\delta^{\mu}_{\nu}\Big )+\mathcal{O}(M^{-6})\nonumber
\end{align}
where $a$, $b_i$'s and $c_i$'s should be read from the equation of motion of $g_{\mu\nu}$ (\ref{eqn-g}). It is straightforward but not easy to show that the final higher order derivative at this level is 
\begin{align}
\mathcal{L}^{(2)}_{HD}=m^2_g\sqrt{-\det g}\Big [\Lambda+c_RR-\frac{c_{RR}}{M^2}\Big(R^{\mu\nu}R_{\mu\nu}-\frac{1}{3}R^{2} \Big )\Big ]+\mathcal{O}(M^{-4})
\end{align}
where
\begin{align}
\Lambda=-2M^2\alpha_0,\qquad\qquad
c_R=1+\alpha^2a^2-\frac{2\alpha_1}{3s_1},\qquad\qquad
c_{RR}=\frac{1}{s_1^2}\Big (2\alpha^2a^2s_1-\alpha_2+\frac{2s_2}{3s_1}\alpha_1\Big ),
\end{align} 
where we have defined
In the following we will use the following definitions to make the relations simpler

\begin{ceqn}
	\begin{equation}
		s_k\equiv\sum_{n=k}^{3}
		\begin{pmatrix}
			3-k\\
			n-k
		\end{pmatrix}a^n\beta_n,\qquad
		\alpha_k\equiv\sum_{n=k}^{4}
		\begin{pmatrix}
			4-k\\
			n-k
		\end{pmatrix}a^n\beta_n,\qquad\qquad k=0,1,2,3.
	\end{equation}
\end{ceqn}
So up to $\mathcal{O}(M^{-4})$ we get a special second order gravity action (\ref{HD-2}) which has conformal gravity term in addition to standard Einstein-Hilbert term in the presence of a cosmological constant.
\subsubsection{PM case}
If we impose PM conditions (\ref{PM-param}) then we will have
\begin{align}
\mathcal{L}^{(2)}_{PM}=-\frac{\alpha^2}{2\beta_2}\frac{m^2_g}{M^2}\sqrt{-\det g}
\Big (R^{\mu\nu}R_{\mu\nu}-\frac{1}{3}R^{2} \Big )+\mathcal{O}(M^{-4})
\end{align}
which is exactly the conformal gravity. This action has a conformal symmetry in addition to diffeomorphism invariance but unfortunately it has a well-studied ghost. However we know the original Lagrangian (\ref{bi-gravity-action}) is ghost free so we claim that if we could go to infinite order gravity then we had a ghost free action for metric $g_{\mu\nu}$. In the next subsections we will try to go to the cubic and quartic orders. We are aware of that any cutting of the recursion procedure may results in a ghost but if we think to our model as an effective theory then going to higher and higher orders means we push the scale of validity of our model higher and higher.

\subsection{Cubic Order}
If we go one step further and assume the following relation for $S$ 
\begin{eqnarray}\label{S3P}
&&S^{\mu}_{\nu}=a\delta^{\mu}_{\nu}+\frac{1}{M^2}\Big (b_1P^{\mu}_{\nu}+b_2[P]\delta^{\mu}_{\nu}\Big )
+\frac{1}{M^4}\Big (c_1(P^2)^{\mu }_{\nu}+c_2[P]P^{\mu}_{\nu}+c_3[P^2]\delta^{\mu}_{\nu}+c_4[P]^2\delta^{\mu}_{\nu}\Big )\nonumber\\
&+&\frac{1}{M^6}\Big (d_1(P^3)^{\mu }_{\nu}+d_2[P](P^2)^{\mu }_{\nu}+d_3[P^2]P^{\mu}_{\nu}+d_4[P]^2P^{\mu}_{\nu}
+d_5[P]^3\delta^{\mu}_{\nu}+d_6[P^3]\delta^{\mu}_{\nu}+d_7[P][P^2]\delta^{\mu}_{\nu}\Big )\nonumber\\&+&\mathcal{O}(M^{-8})
\end{eqnarray}
which is cubic in terms of $P$ then we can show that the cubic order gravity will be read as
\begin{eqnarray}
\mathcal{L}^{(3)}_{HD}&=&\frac{m^2_g\sqrt{-\det g}}{M^4}\Big (n_1R^{\mu\nu}R^{\rho}_{\mu}R_{\nu\rho}+n_2R^{\mu\nu}R_{\mu\nu}R+n_3R^3+n_4R^{\mu\nu}R^{\rho\sigma}R_{\mu\rho\nu\sigma}+n_5R\Box R+n_6R^{\mu\nu}\Box R_{\mu\nu}\Big )\nonumber\\&+&\mathcal{O}(M^{-6})
\end{eqnarray}
where $ \Box=\triangledown^{\gamma}\triangledown_{\gamma} $ and
\begin{eqnarray}
&n_1=\frac{1}{9s_1^5}\Big [9a^2\alpha^2s_1^2(5s_1-2s_2)+2s_1\Big(2s_2(s_1-3s_2+4\alpha_2)+\alpha_2(\alpha_2-s_1)-\alpha_4(s_1-s_2)\Big)-12s_2^2\alpha_2\Big ]\nonumber\\
&\quad n_2=-n_1+\frac{17a^2\alpha^2}{6s_1^2},\quad n_3=\frac{7}{36}n_1-\frac{11a^2\alpha^2}{18s_1^2},\quad n_4=\frac{-4a^2\alpha^2}{s_1^3}(s_1-s_2),\quad n_6=-3n_5=-\frac{a^2\alpha^2}{s_1^2}.\nonumber
\end{eqnarray}

\subsubsection{PM case}
In the case of PM we will have
\begin{eqnarray}
\mathcal{L}^{(3)}_{PM}&=&-\frac{\alpha^4}{\beta_2^2}\frac{m^2_g}{M^4}\sqrt{-\det g}\Big (R^{\mu\nu}R^{\rho}_{\mu}R_{\nu\rho}-\frac{7}{24}R^{\mu\nu}R_{\mu\nu}R+\frac{1}{24}R^3-\frac{1}{2}R^{\mu\nu}R^{\rho\sigma}R_{\mu\rho\nu\sigma}+\frac{1}{12}R\Box R-\frac{1}{4}R^{\mu\nu}\Box R_{\mu\nu}\Big )\nonumber\\&+&\mathcal{O}(M^{-6})
\end{eqnarray}
if we employ (\ref{PM-param}).

\subsection{Quartic Order}
Finally we generalized our calculations for the quartic order by assuming

\begin{eqnarray}
S^{\mu}_{\nu}&=&a\delta^{\mu}_{\nu}+\frac{1}{M^2}\Big (b_1P^{\mu}_{\nu}+b_2[P]\delta^{\mu}_{\nu}\Big )
+\frac{1}{M^4}\Big (c_1(P^2)^{\mu }_{\nu}+c_2[P]P^{\mu}_{\nu}+c_3[P^2]\delta^{\mu}_{\nu}+c_4[P]^2\delta^{\mu}_{\nu}\Big )\nonumber\\
&+&\frac{1}{M^6}\Big (d_1(P^3)^{\mu }_{\nu}+d_2[P](P^2)^{\mu }_{\nu}+d_3[P^2]P^{\mu}_{\nu}+d_4[P]^2P^{\mu}_{\nu}
+d_5[P]^3\delta^{\mu}_{\nu}+d_6[P^3]\delta^{\mu}_{\nu}+d_7[P][P^2]\delta^{\mu}_{\nu}\Big )\nonumber\\
&+&\frac{1}{M^8}\Big (l_1(P^4)^{\mu }_{\nu}+l_2[P](P^3)^{\mu }_{\nu}+l_3[P^2](P^2)^{\mu }_{\nu}+l_4[P]^2(P^2)^{\mu }_{\nu}
+l_5[P^3]P^{\mu}_{\nu}+l_6[P][P^2]P^{\mu}_{\nu}+l_7[P]^3P^{\mu}_{\nu}\nonumber\\
&+&l_8[P^4]\delta^{\mu}_{\nu}
+l_9[P]^4\delta^{\mu}_{\nu}+l_{10}[P^2][P]^2\delta^{\mu}_{\nu}+l_{11}[P][P^3]\delta^{\mu}_{\nu}+l_{12}[P^2]^2\delta^{\mu}_{\nu}\Big )
+\mathcal{O}(M^{-10})
\end{eqnarray}
which results in the quartic order gravity
\begin{eqnarray}
\mathcal{L}^{(4)}_{HD}&=&\frac{m^2_g\sqrt{-\det g}}{M^6}\Big (r_1R^{\mu\nu}R^{\sigma}_{\mu}R^{\rho}_{\nu}R_{\sigma\rho}+r_2R^4+r_3R^{\mu\nu}R_{\mu\nu}R^2+r_4R^{\sigma}_{\mu}R^{\mu\nu}R_{\nu\sigma}R
+r_5R^{\mu\nu}R^{\sigma\rho}R_{\mu\sigma\nu\rho}R\nonumber\\
&+&r_6R^{\mu\nu}R_{\mu\nu}R^{\sigma\rho}R_{\sigma\rho}
+r_7R\triangledown_{\rho}R^{\mu\nu}\triangledown^{\rho}R_{\mu\nu}+r_{8}RR^{\mu\nu}\Box R_{\mu\nu}
+r_{9}R^2\Box R+r_{10}RR_{\mu\nu}\triangledown^{\mu}\triangledown^{\nu}R\nonumber\\
&+&r_{11}R^{\rho}_{\mu}R_{\nu\rho}\triangledown^{\mu}\triangledown^{\nu}R+r_{12}R^{\nu\rho}\triangledown^{\mu}R\triangledown_{\rho}R_{\mu\nu}+r_{13}R\triangledown_{\nu}R_{\mu\rho}\triangledown^{\rho}R^{\mu\nu}
+r_{14}R^{\rho}_{\mu}R^{\mu\nu}\Box R_{\nu\rho}+r_{15}R^{\mu\nu}R^{\rho\sigma}\triangledown_{\sigma}\triangledown_{\rho}R_{\mu\nu}\nonumber\\
&+&r_{16}R^{\mu\nu}R^{\rho\sigma}\triangledown_{\sigma}\triangledown_{\nu}R_{\mu\rho}+r_{17}R^{\mu\nu}\triangledown^{\rho}R_{\nu\sigma}\triangledown^{\sigma}R_{\rho\mu}
+r_{18}R^{\mu\nu}\triangledown_{\nu}R_{\rho\sigma}\triangledown^{\sigma}R^{\rho}_{\mu}\Big )+\mathcal{O}(M^{-8})\nonumber
\end{eqnarray}


\subsubsection{PM case}
PM assumption will set the relative coefficients as
\begin{eqnarray}
\mathcal{L}^{(4)}_{PM}&=&-\frac{\alpha^6}{\beta_2^3}\frac{m^2_g}{M^6}\sqrt{-\det g}\Big (\frac{11}{16}R^{\mu\nu}R^{\sigma}_{\mu}R^{\rho}_{\nu}R_{\sigma\rho}-\frac{13}{576}R^4+\frac{31}{144}R^{\mu\nu}R_{\mu\nu}R^2-\frac{13}{24}R^{\sigma}_{\mu}R^{\mu\nu}R_{\nu\sigma}R\nonumber\\
&+&\frac{1}{48}R^{\mu\nu}R^{\sigma\rho}R_{\mu\sigma\nu\rho}R-\frac{29}{192}R^{\mu\nu}R_{\mu\nu}R^{\sigma\rho}R_{\sigma\rho}
+\frac{1}{24}R\triangledown_{\rho}R^{\mu\nu}\triangledown^{\rho}R_{\mu\nu}-\frac{11}{1152}R^2\Box R-\frac{1}{12}RR^{\mu\nu}\Box R_{\mu\nu}\nonumber\\\nonumber
&-&\frac{1}{144}RR_{\mu\nu}\triangledown^{\mu}\triangledown^{\nu}R+\frac{1}{48}R^{\rho}_{\mu}R_{\nu\rho}\triangledown^{\mu}\triangledown^{\nu}R+\frac{1}{4}R^{\mu\nu}R^{\rho\sigma}\triangledown_{\sigma}\triangledown_{\nu}R_{\mu\rho}+\frac{1}{4}R^{\mu\nu}\triangledown^{\rho}R_{\nu\sigma}\triangledown^{\sigma}R_{\rho\mu}\nonumber\\
&-&\frac{1}{24}R^{\nu\rho}\triangledown^{\mu}R\triangledown_{\rho}R_{\mu\nu}-\frac{1}{16}R\triangledown_{\nu}R_{\mu\rho}\triangledown^{\rho}R^{\mu\nu}-\frac{1}{4}R^{\mu\nu}R^{\rho\sigma}\triangledown_{\sigma}\triangledown_{\rho}R_{\mu\nu}
\Big )+\mathcal{O}(M^{-8}),
\end{eqnarray}
where we have used Bianchi identities and removed the total derivative terms. Note that PM condition makes $r_{14}$ and $r_{18}$ to be zero.

\section{Conclusions}
In this work we generalized the idea in \cite{Hassan:2013pca} by studying higher order terms. We could find the cubic and quartic Lagrangian corresponding to Hassan-Rosen bi-gravity model (\ref{bi-gravity-action}) and for its partially massless case. This idea is interesting since it is well-known that Hassan-Rosen bi-gravity action is ghost free. Consequently, we expect its higher order gravity should be ghost free too but one needs to keep infinite order of Riemann tensor to avoid the ghost. However even studying corresponding $n$th order gravity will be useful if we look at it as an effective field theory. In this viewpoint by adding higher order terms we can push ghost's energy scale to higher energy scales so can trust to our model for a bigger range of energy. The parameter which control the validity of effective theory is graviton mass in the original bi-gravity action (\ref{bi-gravity-action}). If the graviton mass goes to zero then bi-gravity action reduces to standard Einstein-Hilebert action \cite{Akrami:2015qga} and if graviton mass was not negligible then one should work with the exact bi-gravity action which means the corresponding higher order gravity should contain infinite terms. It means the graviton mass controls the validity of perturbative approach.

The idea which has been studied here could be investigated more deeply. One open question is to show explicitly that ghost energy scale will be pushed up by adding higher order terms. In the case of PM for the quadratic order one has an additional symmetry i.e. conformal symmetry. It is a very interesting question to ask if a symmetry (e.g. (generalized) conformal symmetry) exist for cubic and quartic actions. We remain these questions for future works.

Note: During the final preparation of the present paper another paper \cite{Gording:2018not} appeared with similar idea. However the only minor differences are just technical issues. I) In comparison to our relation (\ref{S3P}) they consider a more general case (see (3.7) in their paper). We will generalize our model to make it comparable with their results. II) We did our calculations to the fourth order instead of cubic order with given explicit form of Lagrangian for partially massless case.  

\textit{Acknowledgments:}
We are grateful to Angnis Schmidt-May for fruitful discussions at early stages of this work.  We also acknowledge the use of the
Mathematica package xAct \cite{xAct}.


\end{document}